\documentstyle[12pt,iopconf,epsfig]{article}
\begin{document}

\title{Numerical Models of Newtonian and Post-Newtonian 
Binary Neutron Star Mergers}


\author{Edward Y.M. Wang\dag\ddag, 
        F. Douglas Swesty\S\ddag, and 
        Alan C. Calder\ddag}

\affil{\dag\ Department of Physics, Washington University in St. Louis, 
\& Department of Physics \& Astronomy, State University of New York 
at Stony Brook}
\vspace{-2.0mm}
\affil{\ddag\ National Center for Supercomputing Applications, 
University of Illinois at Urbana-Champaign, Urbana, IL, 61801, USA}
\vspace{-2.0mm}
\affil{\S\ Department of Astronomy, 
University of Illinois at Urbana-Champaign, Urbana, IL, 61801, USA}

\vspace{-2.0mm}
\beginabstract
This article describes a comparison of two calculations of the merger 
of a binary neutron star (NS) system which is initially within the 
tidal instability as described by Rasio \& Shapiro \cite{RS1,RS2}.  
The same initial data is used with one simulation involving a 
purely Newtonian evolution of the Euler equations for compressible 
fluids and another similarly evolved except there is an inclusion 
of a gravitational radiation reaction (GRR) term at the 2.5 
Post-Newtonian (PN) order as prescribed by Blanchet, Damour, \& 
Sch\"{a}fer \cite{BDS}.  The inclusion 
of GRR is to allow an approximation of the full relativistic effect 
which forces the inspiral of binary systems.  The initial data is 
identical at the start of each evolution and only the inclusion of 
the 2.5PN term differs in the evolutions.  We chose two co-rotating, 
$\Gamma$=2 polytropic stars with masses $1.4M_{\odot}$, 
radii $R_{*}$=9.56km and central densities of 
$2.5\times10^{15}g{\Large /}cm^{3}$.  The initial binary separation 
is $2.9R_{*}$, inside the region where the dynamical tidal 
instability has been shown to cause the merger of the two stars.  
Comparisons between measured quantities will indicate a 
substantial difference between the two evolutions.  It will be shown 
that the effect of GRR on the dynamics of the merger is significant.
\endabstract

\vspace{-159.10mm}
\begin{small}
\noindent (This manuscript will appear in the proceedings of the 
Second Oak Ridge Symposium on Atomic and Nuclear Astrophysics.)
\end{small}
\vspace{+147.0mm}

\vspace{-5.0mm}
\section{Introduction}
We study the evolution of binary NS systems as it represents one 
of the most interesting astrophysical systems of the known 
universe.  Since the discovery of the first binary pulsar system 
in 1974 by Hulse \& Taylor \cite{HT} and the subsequent 
measurement of the orbital period decrease as predicted by general 
relativity it is has been thought that the energy loss from these 
systems, in the form of gravitational waves (GWs), could be 
measured.  Such measurements are expected to begin early in the 
next century when several GW observatories or detectors, such as 
the US-based LIGO experiment \cite{LIGO}, are to come 
on-line in their advanced stages.  These systems also represent 
interesting laboratories for nuclear physics.  Most recently, the 
site of $\gamma$-ray bursts has been suggested as binary NS 
mergers\cite{PA86,NPP92}.  See Calder, Swesty \& Wang 1988 for 
discussion \cite{ALAN}.

\section{Method of simulation\label{sec:numerics}}

\subsection{Numerical hydrodynamics and self gravity}
We assume the neutron star can be described by a compressible fluid 
which satisfies the Euler equations, plus the solution to Poisson's 
equation for self-gravity and the 2.5PN GRR potential $\Psi$ which 
enters into the momentum equation (see \cite{BDS,ALAN} 
for a discussion of the the Euler equations with self-gravity).  
The method is similar to the one employed in the ZEUS codes 
\cite{ZEUS}.  Usage of the 2.5PN GRR potential is consistent 
with \cite{BDS} for hydrodynamic evolution up to 2.5PN order.  

\subsection{2.5PN gravitational radiation reaction potential $\Psi$ \& boundary conditions}
Solving for the 2.5PN GRR potential $\Psi$ at any time during the 
evolution requires a solution of a Poisson equation of the form 
\begin{equation}
\nabla ^2 R = 4 \pi G D^{ij} x_i \frac{\partial \rho}{\partial x^j}
\end{equation}
where$D^{ij}$=${d^3}{\skew6\ddot{I\mkern-6.8mu\raise0.3ex\hbox{-}}}^{\:ij}/{dt^3}$ 
is third time-derivative of the symmetric, trace-free mass 
quadrupole tensor, and the 2.5PN GRR potential is defined as 
\begin{equation}
\Psi = \frac{2}{5} \frac{G}{c^5} \left( {R - D^{ij} x_i \frac{\partial \Phi}{\partial x^j}} \right)  
\end{equation}
We utilize a full multi-grid W-cycle algorithm to calculate the 
solution of (1).
To define the boundary values of $R$ we utilize a Riemann sum 
decomposition similar to what is described in \cite{ALAN} for 
the Newtonian potential.  The solution to (1) will 
be of the form, 
\begin{equation}
R = -G \int { \frac{D^{ij} {{x_{i}}^{'}} \frac{\partial \rho}{\partial {x^{j}} '}} { \vert \vec{r}-\vec{r} ' \vert } } dr'
\end{equation}
where $\vec{r}'$ is the distance from the mass element to the 
boundary zone (see \cite{ALAN} for discussion).  We have found that 
$R$=0 boundary conditions are sufficient.

\section{Simulations\label{sec:sims}}

\subsection{Technical concerns: Evolution time scales}
The high densities present in neutron stars implies the adiabatic 
sound speed $C_s$ will be very large.  The Courant-Friedrichs-Lewy 
(CFL) condition defines the minimum time step and since it is 
dependent on the inverse of $C_s$ one expects evolutions on 
physical time scales (milliseconds) to require many thousands of 
numerical time steps.  Explicitly, the CFL condition may be written 
as 
\begin{equation}
{\triangle t} = {\alpha} \left( {\triangle x} {\Large /} C_s \right)
\end{equation}
where ${\alpha}$ is the Courant number.  The time step may also 
be reduced if high velocities are present or if there is 
dissipation \cite{ZEUS}.  A suite of code tests, 
spatial and temporal resolution tests, and questions of stability 
have been addressed \cite{ALAN}.

\subsection{Initial data and atmosphere}
We construct initial data for a close binary system because of the 
computational impracticality of evolving two stars from infinity.  
Initially, two non-rotating, $\Gamma$=2 polytropic stars with 
masses $1.4M_{\odot}$, radii $R_{*}$=9.56 km and central 
densities of $2.5\times10^{15} g{\Large /}cm^{3}$ are placed on a 
grid with an initial binary separation of $2.9R_{*}$ between 
the star centers.  The equation of state we chose is a polytropic 
one which involves a relationship between pressure $P$ and density 
$\rho$, {\it only}.  It should be noted that the equation of state 
for the evolutions is not isentropic but is an ideal gas equation 
of state defined in terms of 
\begin{equation}
P = (\Gamma-1)E = (\Gamma-1)e\rho
\end{equation}
where $e$ is the specific internal energy.  We prescribe a low 
density atmosphere, $\rho$=$1.0{\times}10^{3}$ g{\Large /}{cm$^3$}, 
with a high internal energy.  This prevents the atmosphere from 
falling onto the star surface and shocking over short time scales.  
We have found other treatments of the atmosphere, such as the one 
employed by Ruffert \etal \cite{MAX96}, to be problematic 
and erroneous.  All the fluid elements in the stars and a nearby 
region of atmosphere are then set to co-rotate about the center 
of the grid with a frequency defined by Kepler's third law for 
two point particles.  The axis of rotation is the z-axis.

\subsection{Results}
For the two simulations we consider, the grid size is $129^3$ 
with a $dx$=$0.89$ km allowing for a resolution of $\approx$21 
zones across each star.  The Newtonian evolution, simulation 
N2.9, is evolved for 3776 time steps with a total physical 
time of 5.40ms.  Tidal distortion of the stars begins 
immediately.  By one-eighth of an orbit the surfaces of the 
stars are touching.  After one-half an orbit only the densest 
regions of the stars are not in contact.  However, it is 
after this time during which the cores of the individual 
stars begin to fall inward toward the system center of mass 
that we find mass shedding and angular momentum transfer 
from the cores to the lower density regions.  As these low 
density regions gain angular momentum spiral arms form 
carrying this material away from the coalescing 
object and off the evolution grid.  By seven-eights of an 
orbit the densest parts of the stars are touching and the 
spiral arms are pronounced.  Soon after the peak in 
the gravitational wave luminosity (GWL) occurs the magnitude of 
the GWL and the GWs $h_{+}$ and $h_{\times}$ falls-off 
dramatically, as depicted by $h_{\times}$ in figure 1.  A 
differentially rotating object remains by one and a half 
orbits with the merged core rotating faster than the lower 
density disk surrounding it.  The core is 
non-axisymmetric, as would be indicated by a non-zero 
waveform $h_{\times}$ (figure 1).  

The PN2.9 simulation where GRR at 2.5PN order is included 
is evolved for 3694 time steps with a total physical time 
of 5.26ms.  The results are qualitatively similar to the 
Newtonian simulation except that the merger occurs 
substantially earlier.

The simulations have some substantial differences in the 
gravitational waveforms, the time scales to merger, and the 
equilibrium conserved quantities.  A phase difference 
develops in $h_{\times}$ after 0.8ms due to the inclusion 
of the GRR which accelerates the merger.  By 1.5ms, a 90 
degree phase difference can be seen in $h_{\times}$; the 
merger has occured a quarter orbit earlier.  The mass loss 
that occurs is the means by which the pre-merger stars 
shed angular momentum to form the central object.  
Since GRR reduces the total angular momentum of the system 
throughout the evolution in PN2.9 we see a higher total 
mass of the post-merger object or less mass shedding 
than in the N2.9 calculation.  Both mass and angular 
momentum are less in the post-merger N2.9 object after 
the merger (figure 2).  The inclusion of GRR alters the 
evolution of the system and the structure of the 
post-merger object.  Exclusion of GRR clearly ignores an 
effect which significantly changes the evolution of a 
binary NS system.  This effect is not overwhelmed by the 
Newtonian dynamical tidal instability of \cite{RS1,RS2}.  

At 3.485ms into the evolution and nearly 1.8ms past the 
merger the core region can be defined as having a 
torus-like mass distribution (figure 3).  Outward from 
the center to about 3km the mass density rises along the 
$x$- and $y$-axes then falls off rapidly.  As seen in 
figure 4, most of the mass in the system is contained 
within 27km of the system center of mass.  However, a 
significant fraction of the kinetic energy of the system 
lies outside this region (figure 5).  The merged object 
is differentially rotating (figure 6) with the merged 
core rapidly rotating inside low density disk.  The 
post-merger object appears to be nearly axisymmetric, 
and more so than in N2.9.

In conclusion, the effect of GRR on the evolution of the 
binary NS merger is significant.  There is a substantial 
difference between the Newtonian and Newtonian+2.5PNRR 
evolutions.  The final merged objects are different in 
structure and total angular momentum.  The differing 
pre-merger inspiral rate produces a phase difference in 
the waveforms and the total radiated energy through GWs 
is somewhat different.  Thus, it can be stated that the 
effect of the 2.5PNRR potential is significant in the 
late portion of the merger and in forming the final 
coalesced object.  This further motivates the need for 
fully relativistic simulations.  

\section*{Acknowledgments}
We would like to thank Bruce Fryxell, Mike Norman, and 
the rest of our Neutron Star Merger Grand Challenge 
colleagues for many helpful discussions.  We thank NCSA 
and PSC for computing resources under Metacenter 
allocation \#MCA975011 and funding from NASA under a NASA 
ESS/HPCC CAN, NCCS5-153.

\newpage

\hoffset 0.0in
%
%
\begin{figure}
\vspace{-4.0mm}
\begin{center}
\epsfig{file=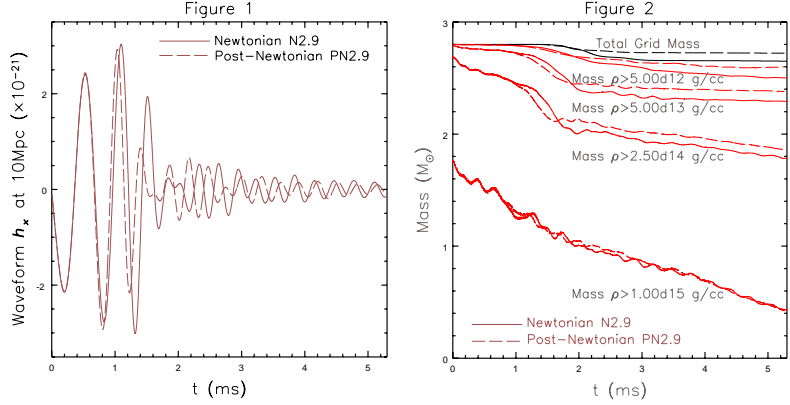,angle=0,height=2.1in }
\end{center}
\vspace{-7.0mm}
\caption{\small Gravitational waveform ${h_{\times}}$ is 
shown for 5.26ms as seen by an observer along the $z$-axis.}
\end{figure}
\begin{figure}
\vspace{-12.0mm}
\caption{\small The mass for a similar time at 
the four density thresholds.}
\end{figure}

%
%
\begin{figure}
\vspace{-8.00mm}
\begin{center}
\epsfig{file=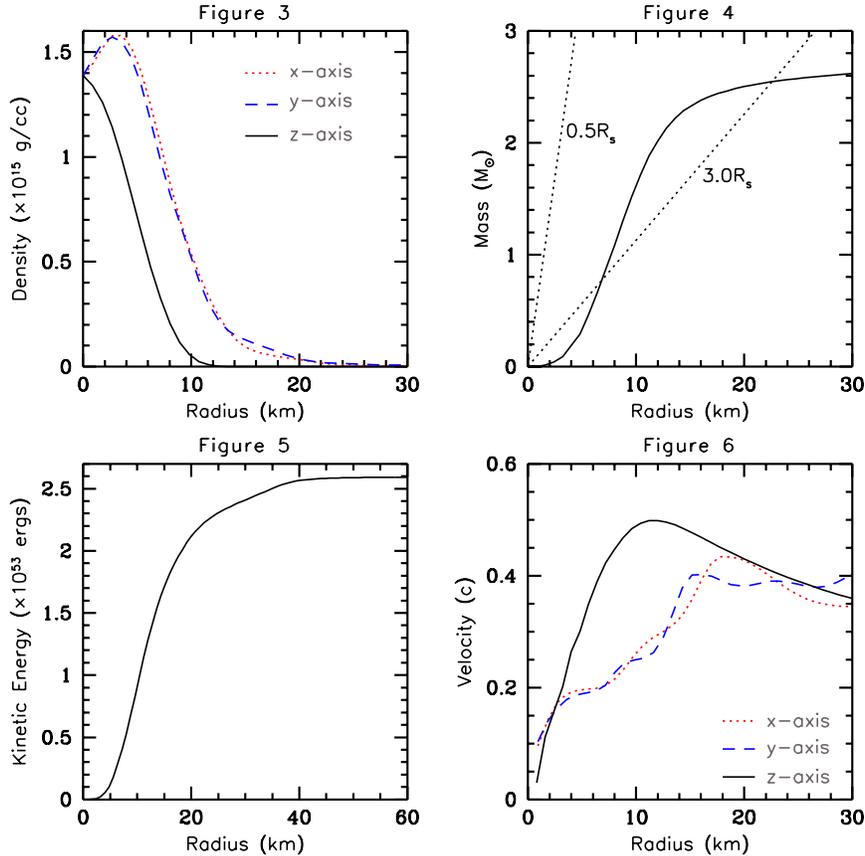,angle=0,height=4.8in}
\end{center}
\vspace{-10.0mm}
\caption{\small Mass density as a function of radius.}
\end{figure}

\begin{figure}
\vspace{-12.0mm}
\caption{\small Total mass enclosed as a function of radius as 
a solid line while the dotted lines are factors of 
the Schwarzschild radius $R_s$.}
\end{figure}

\begin{figure}
\vspace{-12.0mm}
\caption{\small Kinetic energy as a function of radius.}
\end{figure}

\begin{figure}
\vspace{-12.0mm}
\caption{\small The norm of the fluid velocity as a 
function of the radius.}
\end{figure}

\newpage

\end{document}